\documentclass[twocolumn,secnumarabic,amssymb, superscriptaddress, nobibnotes, aps, prb]{revtex4-2}
\usepackage{amssymb,epsfig,graphicx,multirow,bm,dcolumn}

\begin{document}

\title{Structural defects responsible for strain glassy transition in Ni$_{50+x}$Ti$_{50-x}$}

\author{R. Nevgi}
\affiliation{Department of Physics, Goa University, Taleigao Plateau, Goa 403206 India}
\author{Simone Pollastri}
\affiliation{Elettra-Sincrotrone Trieste S.C.p.A., Strada Statale 14, km 163.5, I-34149 Basovizza, Trieste, Italy}
\author{Giuliana Aquilanti}
\affiliation{Elettra-Sincrotrone Trieste S.C.p.A., Strada Statale 14, km 163.5, I-34149 Basovizza, Trieste, Italy}
\author{K. R. Priolkar}
\email{krp@unigoa.ac.in}
\affiliation{Department of Physics, Goa University, Taleigao Plateau, Goa 403206 India}
\date{\today}

\begin{abstract}
The strain glassy phase is produced by doping a small percentage of impurity in a martensitic alloy. Its ground state is conceived to consist of martensitic nano domains spatially separated from each other by a defect phase. The present study, by probing the local structure around the Ni and Ti in martensitic and strain glassy compositions of Ni$_{50+x}$Ti$_{50-x}$, for the first time, identifies the defect phase that is responsible for inhibiting the long range ordering of the elastic strain vector leading to the formation of the strain glassy phase.
\end{abstract}

\maketitle

\section{Introduction}

Shape memory alloys are an important class of functional materials by virtue of their shape memory effect and superelastic properties \cite{Shimizu1987, Wayman1998}. These properties emanate from the martensitic transformation, which is a diffusion less transition wherein the motion of the atoms is consistent with symmetry of the crystal \cite{Tegus4152002,Giot992007,Yu932003}. The atomic displacement below the transition temperature T$_M$ is responsible for changes in the physical properties of the alloys. Hence, the structural changes especially at the atomic level mark an important step in understanding the phenomenon of martensitic transformation \cite{Huang22003,Bhobe742006,Righi552007,Devi972018}. As the motion of atoms is supposed to be consistent with the symmetry of the crystal structure, the martensitic state is classified with a long range ordering of the elastic strain vector along a certain crystallographic direction. However, recent studies have shown that the martensitic transition in shape memory alloys can also occur via a strain glassy phase \cite{Sarkar952005,Lloveras1002008,Vasseur812010}.  Strain glass state is defined as a frozen disordered lattice state procured from dynamically disordered strain state seen in defect containing ferroelastic/martensitic systems. It exhibits a frequency dependent anomaly described by the existence of a dip in the ac storage modulus curve and corresponding peak in loss ($\tan\delta$) curve at glass transition temperature $T_g$ obeying the Vogel-Fulcher law. The strain glass phase is also characterized by the ergodicity breaking evidenced in zero field cooled (ZFC) and field cooled (FC) experiments. The average crystal structure remains invariant across the glass transition accompanied by nano sized domains with frozen elastic strain vector.\cite{Ren2012} Some of the examples of shape memory alloys undergoing strain glass transition are listed in the table \ref{table:table1}.

\begin{table}[htbp]
\caption{\label{table:table1} Critical concentration $x_c$ of the dopants for strain glass transitions.}
\begin{ruledtabular}
\begin{tabular}{cccc}
\textbf{Martensite} & \textbf{Dopant} & \textbf{$x_c$} & \textbf{References}\\[0.5ex]
\hline\hline
Ni$_{50+x}$Ti$_{50-x}$        & Ni         & $x \geq 2$ & \cite{Sarkar952005}\\

Ti$_{50}$Ni$_{50-x}$D$_{x}$   & D = Fe     & $x \geq 6$ & \cite{Wang201058}\\
                              & D = Co     & $x \geq 9$ & \cite{Zhou201058}\\
                              & D = Cr     & $x \geq 4.5$ & \\
                              & D = Mn     & $x \geq 5.5$&\\

Ti$_{50}$Pd$_{50-x}$D$_{x}$    & D = Cr         & $x \geq 9$ &\cite{Zhou200995}\\
                               & D = Fe         & $x > 14$ &\cite{Zhou2014251}\\
                               & D = Mn         & $x > 13$ &\\

Ni$_{55-x}$Co$_{x}$Fe$_{18}$Ga$_{27}$ & Co     & $x \geq 10$ &\cite{Wang201298}\\

Ni$_{55-x}$Co$_{x}$Mn$_{20}$Ga$_{25}$ & Co     & $x \geq 10$ &\cite{Wang2012101}\\

Ni$_{50}$Mn$_{37.5-x}$Fe$_{x}$In$_{12.5}$ & Fe & $x \geq 2.5$ &\cite{Nevgi2018112}\\[1ex]
\end{tabular}
\end{ruledtabular}
\end{table}

The data presented in table \ref{table:table1} suggest that a dopant concentration of $\le$ 10 at.\% induces the strain glassy state in a martensitic alloy.  The strain glassy state proceeds via percolation \cite{Zong2019123} when a minor fraction of impurity dopant is added to a ferroelastic material. Similar observations can be made about the relaxor ferroelectric materials and the cluster spin glassy materials \cite{Viehland1991120,Shvartsman2009379,Cui2010107,Blascol201224,Djurberg199779,Kumar201395} wherein a small amount of impurity dopant forces a non-ergodic ground state in a ferroelectric or ferromagnetic material. This situation is in contrast to canonical spin glasses. A canonical spin glass state results when a few magnetic atoms are dissolved in a non magnetic matrix.  For instance, in AuFe spin glass, the concentration of magnetic Fe is only about 25\%, and the nonmagnetic impurity is 75\% \cite{Cannella19726}. The structural modifications caused by the minority impurity atoms in the ferroic materials are yet to be understood. In particular, one needs to identify the structural defects induced by the doped atoms/ions to precipitate a non-ergodic ground state in such ferroic materials.

The binary NiTi is a prototypical martensitic alloy in which, the studies have shown that, any impurity addition including Ni itself drives the alloy towards a strain glassy transition (see Table \ref{table:table1}). Interestingly, the addition of just $\sim$ 2\% of excess Ni impurity disturbs the long range ordering of the elastic strain vector \cite{Sarkar952005}. Such a small concentration of the dopant atoms further necessitates the identification of structural defects that prevent the long range ordering  of the elastic strain vector. With the crystal structure remaining invariant, a decisive knowledge of the evolution of short range structural order is needed. The \emph{in-situ} high resolution electron microscopy (HREM) studies on Cr doped TiPd alloys have shown that the ground state is an inhomogeneous distribution of nano sized domains with modulated structure evidenced from incommensurate reflections below glass transition temperature \cite{Zhou1122014}. The \emph{in-situ} anomalous small-angle X-ray scattering (ASAXS) technique on Ni$_{51.3}$Ti$_{48.7}$ has revealed the crucial role of Ni atoms as point defects in the mechanism of strain glass transition. It has been shown that the Ni atoms are distributed in nano domains consisting of disk-like core--shell configuration with a Ni-rich shell and a highly Ni-rich core \cite{Huang102020}. However, the nature of structural defects and their ability to keep the crystal structure invariant despite the presence of martensitic domains is still not clearly understood. Extended x-ray absorption fine structure (EXAFS) is one of the best techniques to map the local structural distortions in the vicinity of the absorbing atom due to the addition of the impurity. In this letter, we report EXAFS study at the Ni and Ti K edges in Ni$_{50+x}$Ti$_{50-x}$, $0 \le x \le 10$, alloys with the aim of understanding the structural defects and the interactions responsible for inhibiting the long range ordering of the elastic strain vector leading to the realization of a glassy state. The EXAFS data analysis reveals a formation of BCC Ni clusters in the Ni excess NiTi alloys. These Ni clusters are responsible for inducing the strain glassy phase in the $x = 2$ alloy.

\section{Experimental techniques}

The alloys Ni$_{50+x}$Ti$_{50-x}$ ($x$ = 0, 2, 5, 10 identified as NT50, NT52, NT55 and NT60 respectively ) were prepared by arc melting technique in Argon atmosphere followed by annealing at 1273K for 1 hour and quenching in ice cold water. Scanning electron microscopy with energy dispersive x-ray (SEM-EDX) measurements were performed on the alloys so formed and the compositions so obtained were found to be within 3$\%$ of stoichiometric values. X-ray diffraction patterns were recorded on powdered samples using Cu K$_\alpha$ radiation in the angular range of 20$^\circ$ to 100$^\circ$ and phases were identified by LeBail refinement using Jana 2006 software\cite{Petricek2292014}. Martensitic transition temperatures were ascertained using differential scanning calorimetry (DSC)  performed using Shimadzu DSC-60 on 7-8 mg pieces of each alloy crimped in aluminium pan. The frequency-dependent measurements of AC storage modulus and internal friction (tan$\delta$) were obtained using Dynamical mechanical analyzer (Q800, TA Instruments) wherein the measurements were performed using 3 point bending mode by applying a small AC stress that generated a maximum displacement of 5 $\mu$m at different frequencies in the range of 1 Hz to 10 Hz on rectangular pieces of (10mm $\times$ 5mm $\times$ 1mm) dimensions. The Extended Xray Absorption Fine Structure (EXAFS) scans were recorded at 300K and at 77K in the transmission mode at XAFS  beamline at Elettra Synchrotron Source using silicon (111) monochromator \cite{Cicco1902009}. The incident and the transmitted photon energies were simultaneously recorded at Ni K edge (8333 eV) and Ti K edge (4966 eV) using gas ionization chambers as detectors. Absorbers were prepared by stacking layers of scotch tape containing uniformly spread powdered alloys and the thickness of the absorbers were adjusted to obtain absorption edge jump $\Delta\mu$ (t) $\leq$ 1 wherein $\Delta\mu$ is the change in absorption coefficient at the absorption edge and t is the thickness of the absorber. The theoretical calculations were computed with ATOMS and FEFF8.4 programs (\cite{Rehr102009}), and the EXAFS data analysis were carried out using well established procedures in the Demeter software \cite{Raval200512}.

\section{Results}
\subsection{Characterization}

\begin{table}[h]
\caption{\label{tab:table2} The alloy codes, nominal and the actual compositions as obtained from the SEM-EDEX measurements within the error bar of 3\% for the four alloy compositions.}
\begin{ruledtabular}
\begin{tabular}{ccc}
\textbf{Alloy code} & \textbf{Nominal}           & \textbf{Actual} \\[0.5ex]
                    &  \textbf{compositions}     &  \textbf{compositions}\\
\hline\hline
NT50 & Ni$_{50}$Ti$_{50}$  & Ni$_{52.06}$Ti$_{47.96}$ \\
\hline
NT52 & Ni$_{52}$Ti$_{48}$  & Ni$_{54.82}$Ti$_{45.18}$ \\
\hline
NT55 & Ni$_{55}$Ti$_{45}$  & Ni$_{57.53}$Ti$_{42.47}$ \\
\hline
NT60 & Ni$_{60}$Ti$_{40}$  & Ni$_{60.61}$Ti$_{39.39}$ \\[1ex]
\end{tabular}
\end{ruledtabular}
\end{table}

The table \ref{tab:table2} provides the actual composition of the four alloys as obtained from the SEM-EDEX measurements and the alloy codes corresponding to each compositions which will be henceforth used in the manuscript.

\begin{figure}[h]
\begin{center}
\includegraphics[width=\columnwidth]{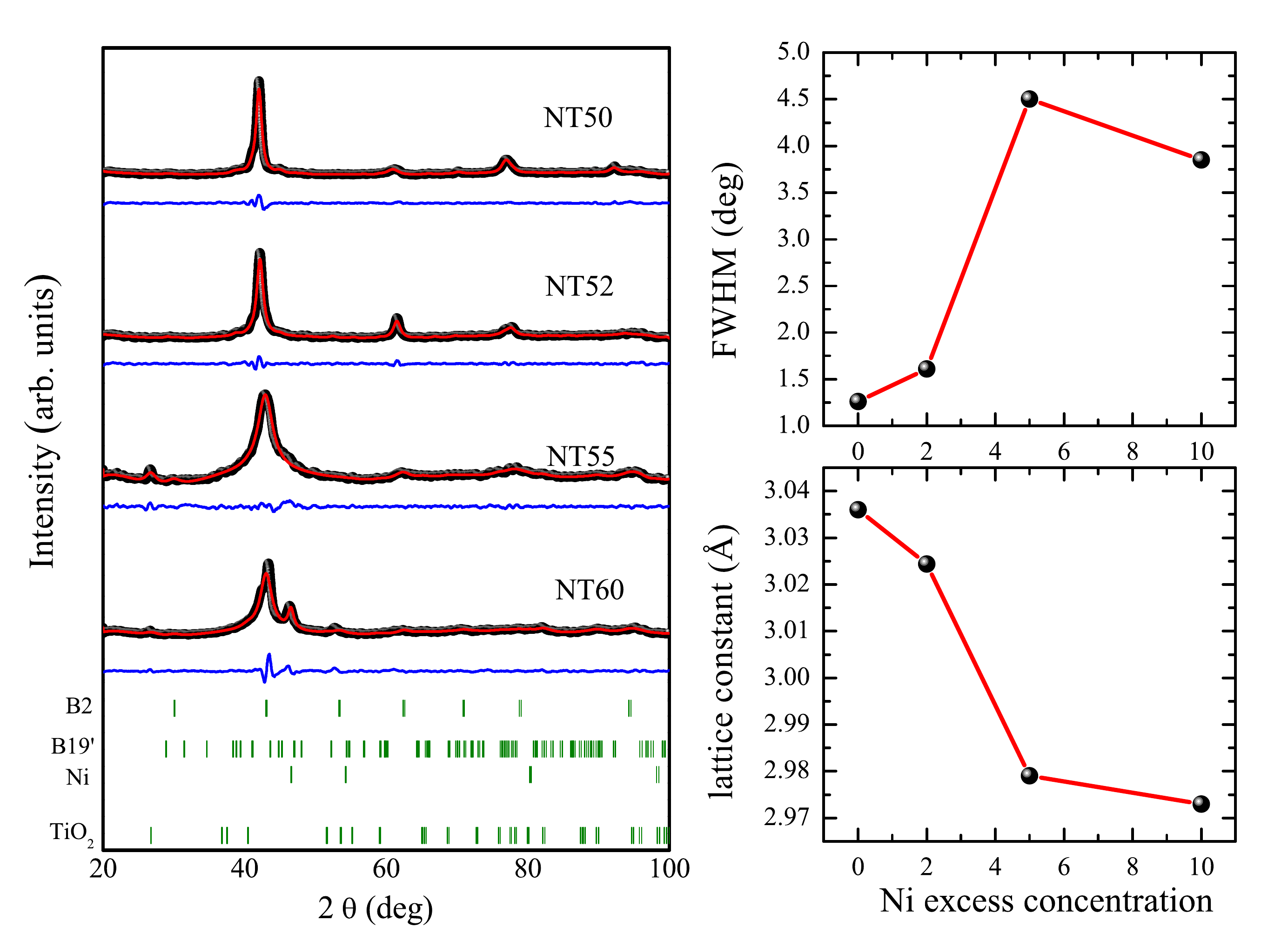}
\caption{LeBail refined x ray diffraction patterns at room temperature for all the four alloy compositions. The figure also indicates a built up of strain explained by the behavior of full width half maximum of the (101) Bragg peak and the decrease in lattice constant with increase in $x$ in Ni$_{50+x}$Ti$_{50-x}$.}
\label{fig:XRD1}
\end{center}
\end{figure}

The LeBail refined x-ray diffraction data recorded at 300 K on the powdered samples is presented in Fig.\ref{fig:XRD1}, showing the major B2 phase in all the four alloy compositions. A small fraction ($\sim$ 9\%) of the B19$'$ phase is noticed in the NT50 alloy, while the alloy NT55 exhibit the B2 phase with broad diffraction peaks with about 6\% anatase TiO$_2$. Further, the alloy NT60 exhibits a segregation of around 20\% impurity face centered cubic (FCC) Ni metal phase in addition to the NiTi B2 phase. The variation of the most intense (101) Bragg reflection of the B2 phase presented in Fig. \ref{fig:XRD1} indicates a decrease in lattice constant with increasing Ni content. The full width at half maximum (FWHM) of the (101) Bragg peak also increases with $0 \le x \le 5$ and decreases slightly in NT60. The systematic increase in the peak's width manifests a build-up of strain that finally results in a phase separation into cubic B2 phase and FCC Ni metal phase accompanied by a reduction in the FWHM of the Bragg peaks in NT60.

\begin{figure}[h]
\begin{center}
\includegraphics[width=\columnwidth]{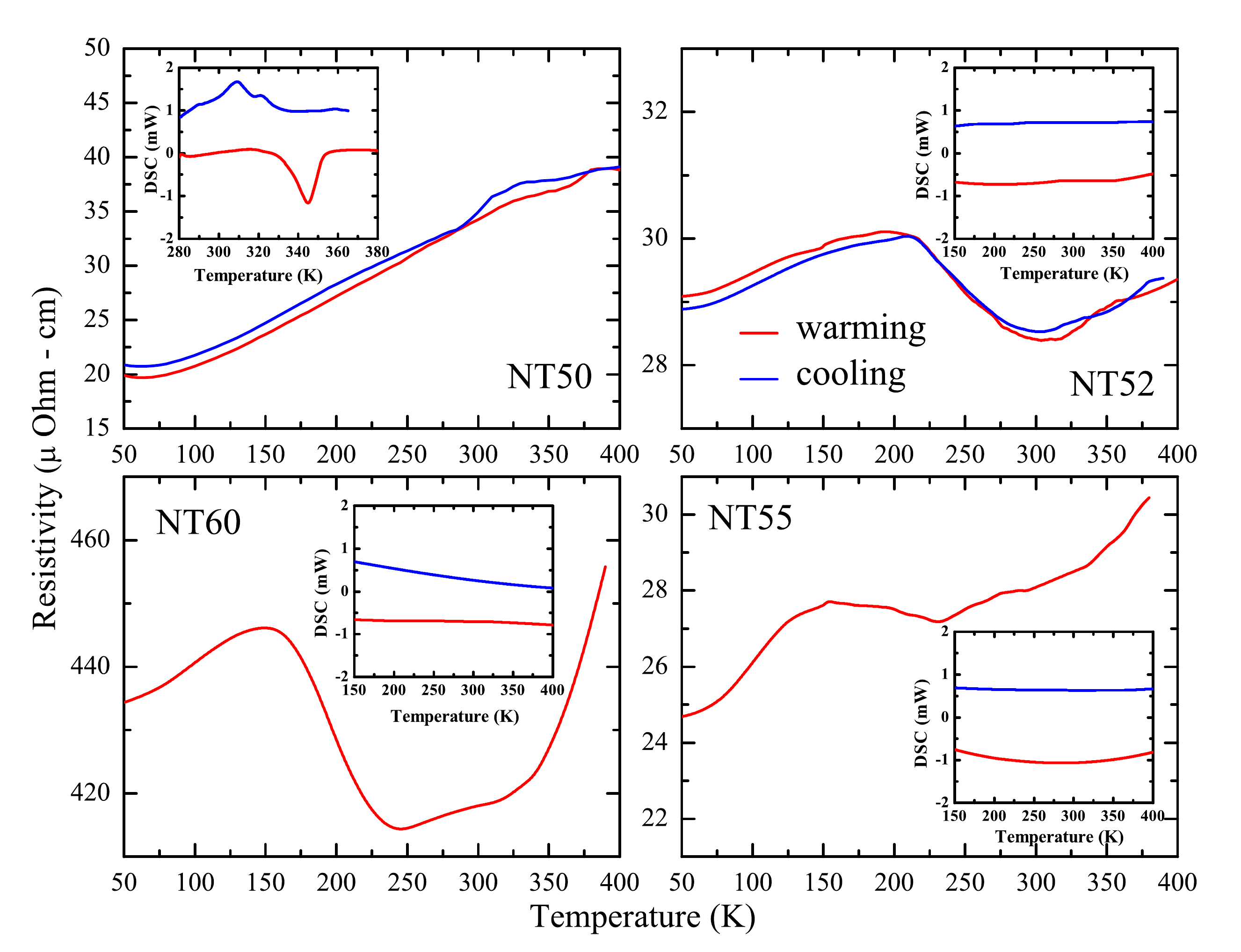}
\caption{The resistivity and the dsc (shown as insets) plots showing martensitic transition in the alloy NT50 just above 300 K and is not seen in the other alloys in the temperature range of 50 K to 400 K.}
\label{fig:DSCRES}
\end{center}
\end{figure}

The alloy NT50 exhibits a martensitic transition with a martensitic start temperature $M_s$ = 330 K, as evidenced from the differential scanning calorimetry (DSC) and resistivity measurements Fig.\ref{fig:DSCRES}. These results are in agreement with the ones reported in \cite{Sarkar952005}.  The transformation hysteresis region, which is defined as the region between martensitic finish temperature $M_f = 280 K$ and austenitic finish temperature $A_f = 350 K$ could be the reason for the  relatively broad diffraction peaks and presence of B19$'$ phase seen in the x-ray diffraction pattern of NT50 at room temperature ($\approx$ 300 K). Martensitic transition is not seen in the DSC measurements recorded down to 150 K or in the resistivity measurements recorded down to 50 K in NT52, NT55 and NT60 alloys. The resistivity of these three alloys instead displays a broad hump below 200 K, which appears at lower and lower temperatures with increasing Ni concentration. An increase in the value of resistivity with Ni excess concentration more evidently visible in NT60 and could be due to increased scattering as a result of impurity doping.

\begin{figure}[h]
\begin{center}
\includegraphics[width=\columnwidth]{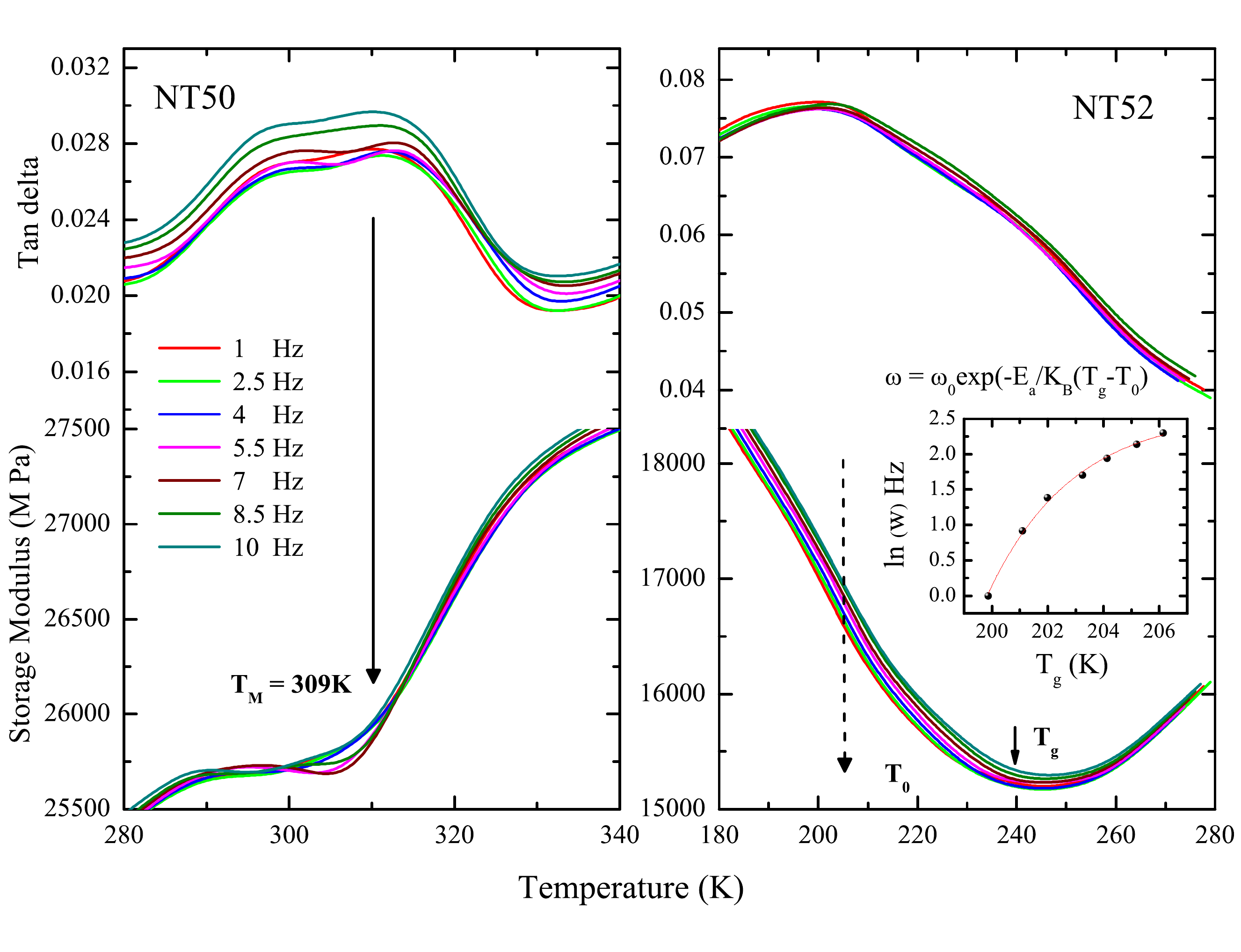}
\caption{The variation of storage modulus and loss as function of temperature at different frequencies in the alloys NT50 and NT52. The frequency dependence at glass transition in NT52 is exhibited by the behaviour of Vogel Fulcher plot seen as an inset. }
\label{fig:DMA}
\end{center}
\end{figure}

Fig.\ref{fig:DMA} shows the variation of the storage modulus and loss (tan $\delta$ ) with the temperature at different frequencies in the alloys NT50 and NT52. In NT50, a dip in storage modulus and a peak in loss signify the martensitic transition. The temperature of the dip and the peak in the two respective signals lie within the martensitic start (M$_S$) and finish (M$_F$) temperature range obtained from the transport property measurements. In NT52, broad humps are noticed in the storage modulus and loss signals at around 200 K, which display a frequency dependence according to the Vogel-Fulcher law. Such a behavior confirms the strain glass transition in the alloy. The DMA measurements on the other two compositions NT55 and NT60 did not show any features down to 150 K which corresponds to the lower data limit of the instrument.

\subsection{EXAFS calculations}

The results so far have shown that with the average structure remaining invariant, the martensitic transition is suppressed by glassy behavior at $x \ge 2$, which propels the investigation of the local structure to understand the interatomic interactions driving the glassy transition. Local structural studies were performed using EXAFS data at the Ni and Ti K edges. Both the EXAFS data were first fitted together with a common structural model based on their crystal structure. This approach gave a reasonably good fit only for NT50, but discrepancies were noticed in all other alloys. Therefore, to ascertain a correct structural model for all the investigated alloys, FEFF 8.4 \cite{Rehr102009} was employed to \textit{ab-inito} calculate EXAFS signals at the Ni K and Ti K edges for the austenitic (B2), the martensitic (B19$'$) and the R phases of NiTi as well as for phases like Ni$_3$Ti and pure Ni metal. These calculated spectra, $\chi(k)$ were respectively compared with the experimental data recorded at 77 K in all four alloys.

\begin{figure}[h]
\begin{center}
\includegraphics[width=\columnwidth]{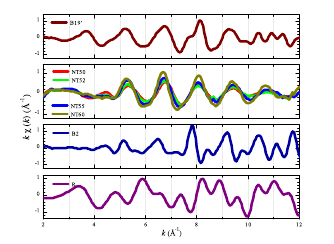}
\caption{Comparison between the $k$-weighted EXAFS spectra at Ti edge of the four alloy compositions recorded at 77 K with the theoretically calculated spectra of the three structures B19$'$, B2 and R showing a favorable resemblance to B19$'$ phase over B2 or R phases.}
\label{fig:cal_Ti}
\end{center}
\end{figure}

Fig.\ref{fig:cal_Ti} displays the $k$-weighted EXAFS spectra recorded at 77 K for the four alloy compositions in the $k$ range of 2 to 12 \AA$^{-1}$, along with the calculated spectra for B2, R and B19$'$ structures. EXAFS oscillations in all four alloys are similar but increase in magnitude with increasing Ni concentration. A good resemblance is noticed between the experimental spectra and the calculated spectra obtained for B19$'$ structure rather than the B2 or R structure. This is expected for the  NT50 alloy as it is in its martensitic state at 77 K. However, for the other Ni excess alloys, similarities between experimental and the calculated spectra for B19$'$ structure indicates modulated local structure around Ti in these alloys.

\begin{figure}[h]
\begin{center}
\includegraphics[width=\columnwidth]{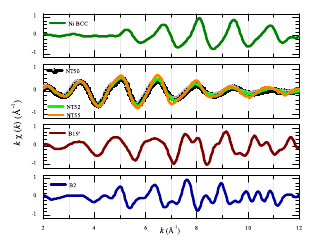}
\caption{Comparison of the $k$-weighted EXAFS spectra at the Ni edge of the alloys NT50, NT52 and NT55 at 77 K with the B2 and B19$'$ phases of the NiTi alloy and Ni BCC metal phase indicating that the Ni local structure can be best described as a combination of B19$'$ phase and Ni BCC in all the three alloy compositions.}
\label{fig:cal_Ni}
\end{center}
\end{figure}

Fig.\ref{fig:cal_Ni}(a) shows the $k$-weighted EXAFS spectra for the NT50, NT52 and NT55. The three alloys display rather analogous EXAFS oscillations, with their amplitudes increasing with excess Ni concentration. A comparison of experimental spectra with the one calculated for B19$'$ structure reveals an apparent discrepancy in the EXAFS oscillations, especially above 7 \AA$^{-1}$. In the region, $k >$ 6 \AA$^{-1}$, some similarities can be seen in the experimental spectra and  the calculated spectra for body centered (BCC) Ni structure \cite{Tian200594} (Fig.\ref{fig:cal_Ni}(b)). In the B2 or B19$'$ structures of NiTi, the Ni atom is surrounded by 8 Ti atoms in the first coordination while in the BCC Ni phase, the nearest neighbors are Ni atoms themselves. This change in the nearest neighbor atom from Ti (Z = 22) to Ni (Z = 28) shifts the back-scattering amplitude maximum from $\sim$ 5.5 \AA$^{-1}$ to about 7 \AA$^{-1}$ \cite{McKale121988,Rehr722000}. Therefore, the similarities of experimental spectral features with BCC Ni calculated spectra implies that the Ni local structure in the three alloys with excess Ni to be a combination of B2/B19$'$ and Ni BCC phase. In these alloys, the Ti EXAFS matches with the B19$'$ calculated spectra while Ni EXAFS is equivalent to the sum of B19$'$ and BCC Ni calculated spectra.

\begin{figure}[h]
\begin{center}
\includegraphics[width=\columnwidth]{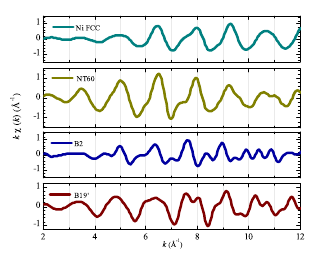}
\caption{Comparison of the $k$-weighted EXAFS spectra at the Ni edge of NT60 at 77 K with the B2 and B19$'$ phases of the NiTi alloy and Ni FCC metal phase indicating a combination of B2 phase and Ni FCC metal phase to be favorable solution.}
\label{fig:cal_NT40}
\end{center}
\end{figure}

The situation is entirely different in NT60. The x-ray diffraction pattern of this alloy showed the presence of FCC Ni in addition to NiTi phase. Therefore, the contribution of FCC Ni to the Ni EXAFS in this alloy is inevitable. This is seen in Fig. \ref{fig:cal_NT40}, wherein the calculated EXAFS at Ni K edge in FCC Ni, B2 and B19$'$ structures of NiTi are compared with experimental EXAFS in Ni$_{60}$Ti$_{40}$. Another interesting point to note is that the Ni local structure in NT60 carries an essence of the B2 phase, unlike the other three alloy compositions that had more similarities with the B19$'$ phase. Hence in these Ni$_{50+x}$Ti$_{50-x}$ alloys, the Ni local structure is different from that of Ti and evolves from a sum of B19$'$ and BCC Ni phases to a mixture of B2 and FCC Ni phases with increasing $x$. This difference in local structures explains the above mentioned unsuccessful attempt of fitting the Ni and Ti EXAFS together.

\subsection{EXAFS fittings}

In order to estimate the near neighbor bond distances of Ni and Ti in these Ni$_{50+x}$Ti$_{50-x}$ alloys, the EXAFS spectra at the two edges were fitted using structural models compiled based on the above comparison. The Ti and Ni edge EXAFS spectra in the $k$ range of 3 to 12 \AA$^{-1}$ and $R$ range of 1 to 3 \AA~ were fitted using a total of nine independent parameters consisting of correction to the bond length $\Delta R$ and mean square variation in bond length $\sigma^{2}$ for each of the scattering paths used. The coordination numbers of the individual scattering paths were obtained from the structural model derived from the reported crystal structure \cite{Sitepu200274,Tian200594,Owen193621}. The amplitude reduction factor, $S_0{^2}$ and correction to edge energy $\Delta E_0$, were obtained from the analysis of the standard metal spectra. These parameters were kept fixed throughout the analysis. In addition, during the analysis of Ni EXAFS spectra, one more independent parameter $y$ was introduced to estimate the fraction of the two phases, NiTi and Ni metal phase. In the case of NT50, $y$ = 0.05 $\pm$ 0.02 and it increased with increasing Ni concentration.

\begin{figure}[h]
\begin{center}
\includegraphics[width=\columnwidth]{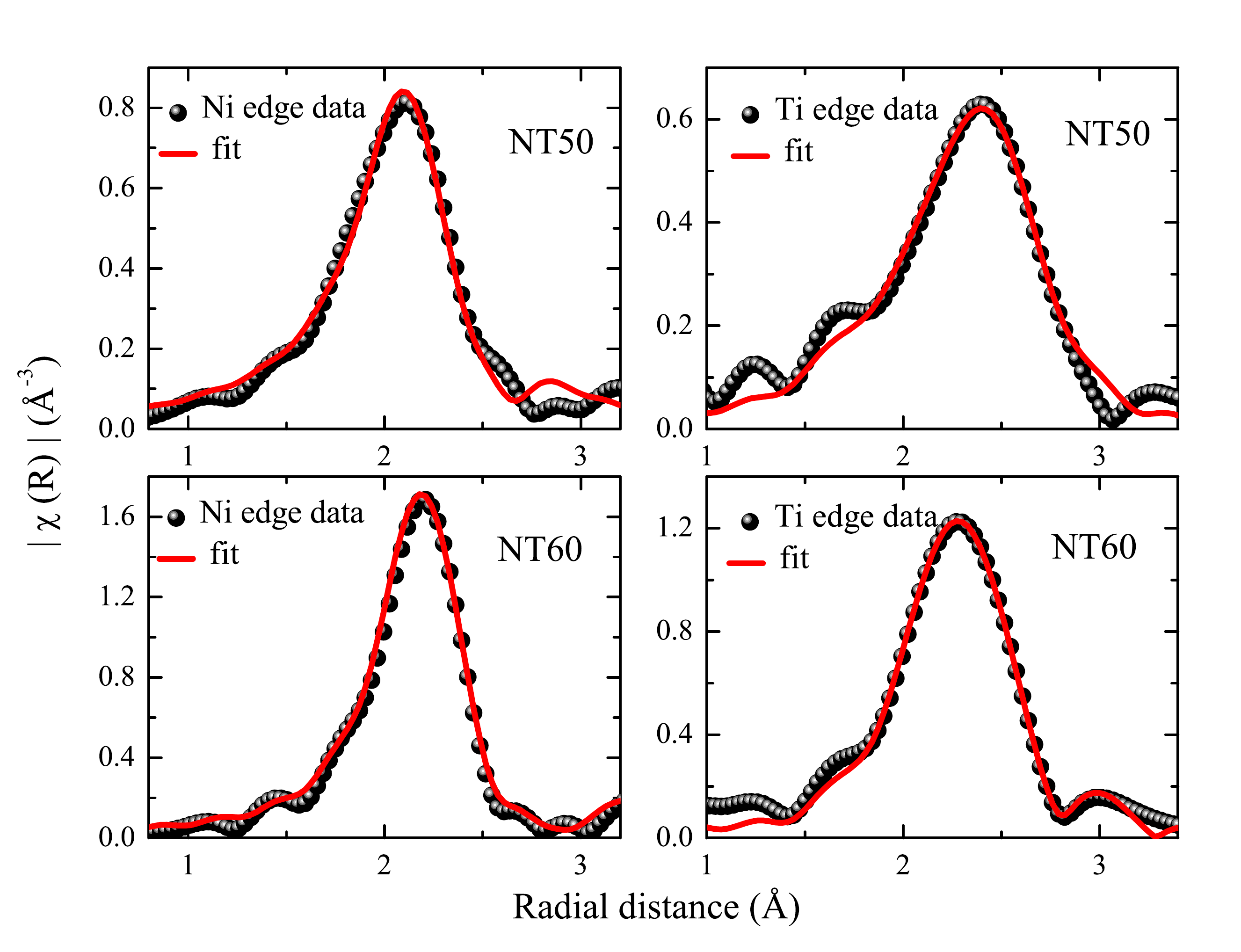}
\caption{The magnitude of Fourier transform of EXAFS spectra at the Ni and Ti edge along with the respective best fits at 77 K for the alloys NT50 and NT60.}
\label{fig:exp_NT50}
\end{center}
\end{figure}

\begin{figure}[h]
\begin{center}
\includegraphics[width=\columnwidth]{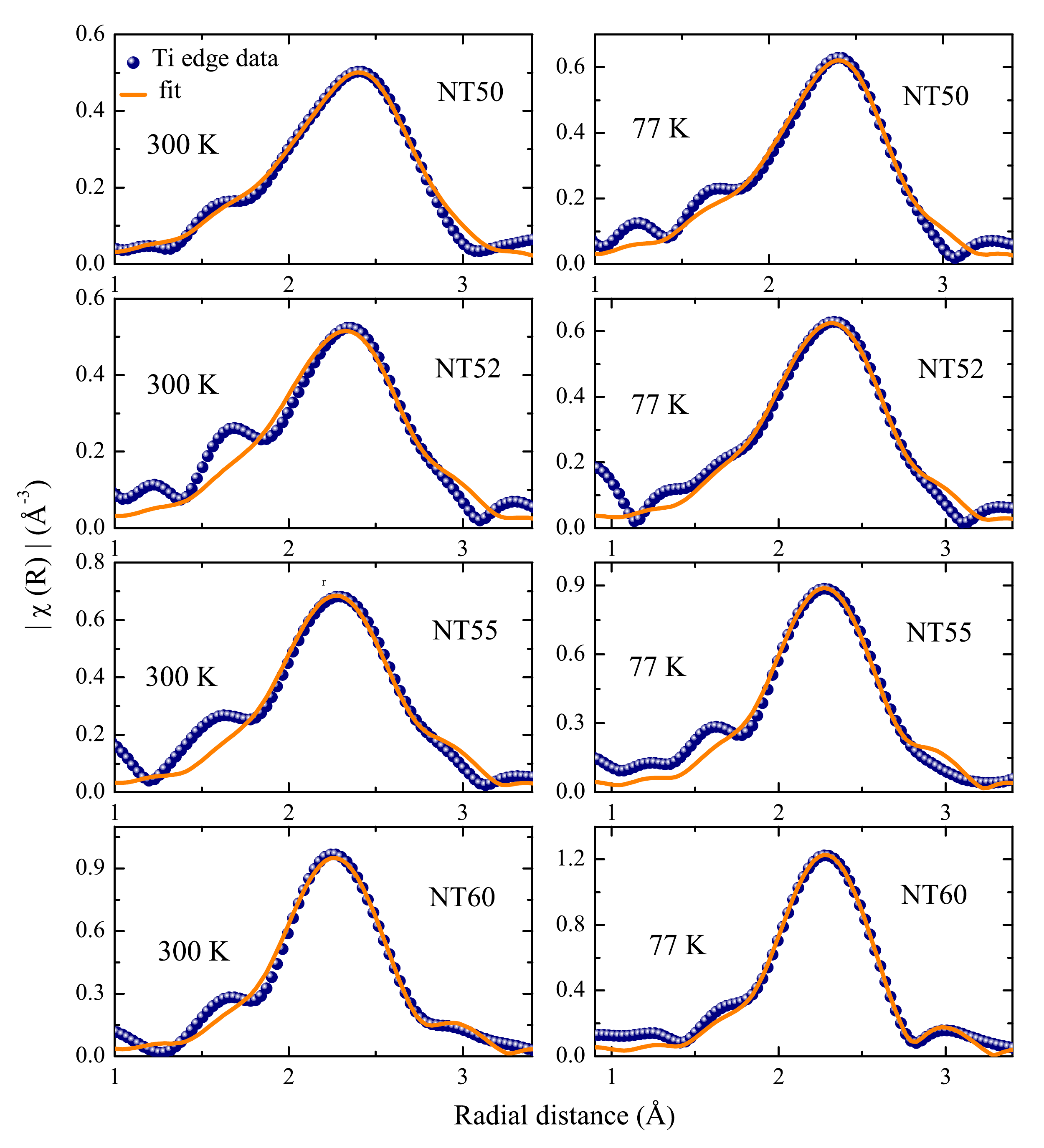}
\caption{The magnitude of Fourier transform spectra at Ti edge for the four alloy compositions at 300K and 77K.}
\label{fig:exp_Ti}
\end{center}
\end{figure}

\begin{figure}[h]
\begin{center}
\includegraphics[width=\columnwidth]{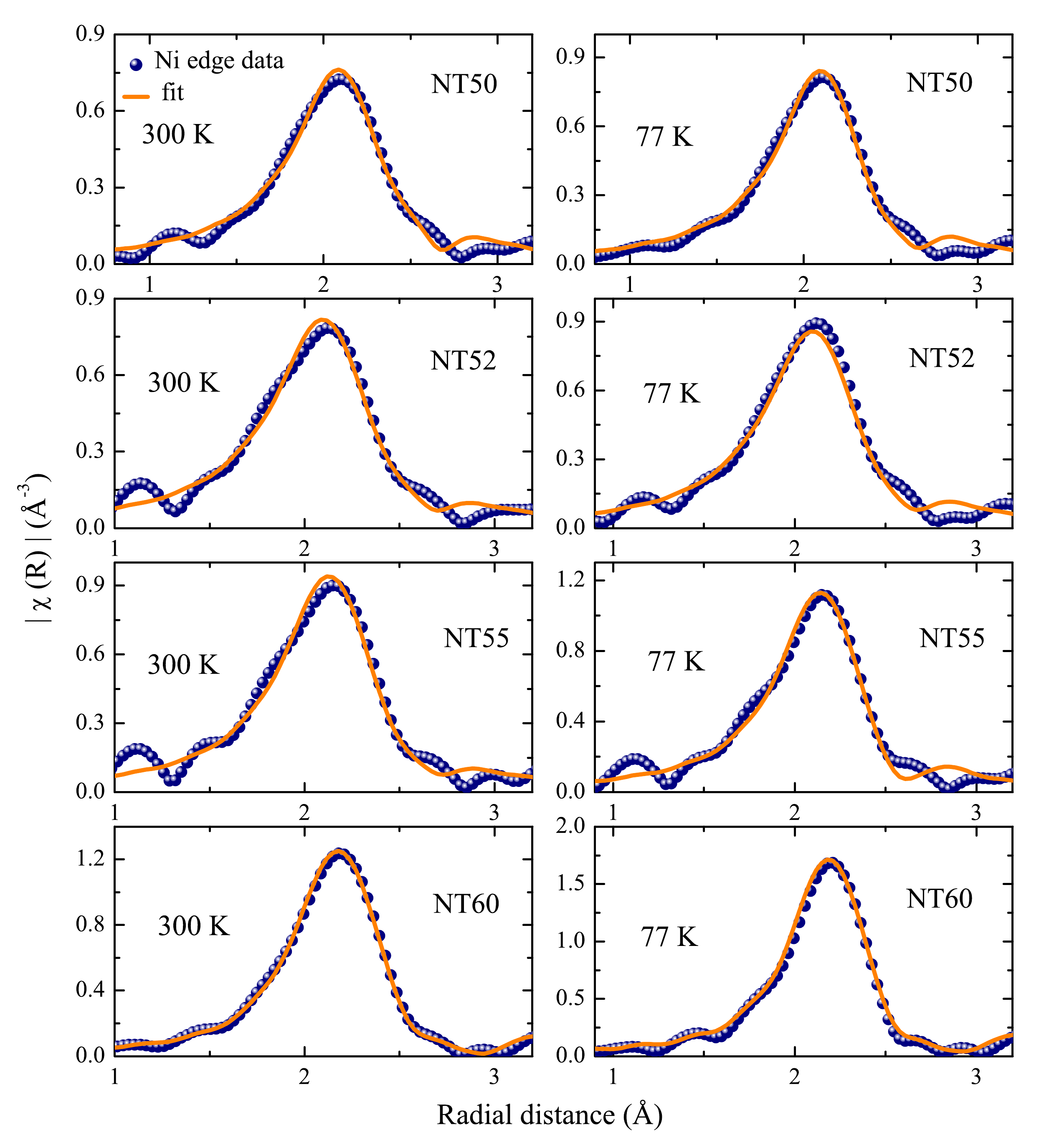}
\caption{The magnitude of Fourier transform spectra at Ni edge for the four alloy compositions at 300K and 77K.}
\label{fig:exp_Ni}
\end{center}
\end{figure}

The Ti local structure in all the four alloys was fitted to correlations arising out of B19$'$ phase. The magnitude of Fourier transform (FT) of the Ti EXAFS spectra over $R$ = 1 \AA~ to 3 \AA~ includes contributions from four coordination shells, two with Ni atoms at $\sim$ 2.52 \AA~ and 2.60 \AA~ and two with Ti atoms at $\sim$ 2.88 \AA~ and 2.99 \AA~ respectively. Fig.\ref{fig:exp_NT50} presents the fittings to the EXAFS spectra in the $R$-space for the alloys NT50  and NT60 at 77 K. The experimental EXAFS data along with best fits at 300 K and 77 K is presented in Fig.\ref{fig:exp_Ti}, while the values of the bond distances and the mean square displacements in the four near neighbor bond are presented in the supplementary material \cite{Supplementary}. We observe no significant changes in the Ti--Ni and Ti--Ti bond distances either with temperature or with a change in Ni concentration. The fittings confirm the local structure of Ti to be like that of B19$'$ phase and does not change with Ni doping in the alloys Ni$_{50+x}$Ti$_{50-x}$.

As mentioned earlier, the Ni local structure comprises of contributions from more than one phase, and hence the EXAFS spectra in NT50, NT52 and NT55 were fitted with two different combinations. The first one consisted of correlations from B2 and BCC Ni phases, and the second one had correlations from B19$'$ and BCC Ni phases. The Ni EXAFS in NT60 was again fitted with two combinations consisting of correlations from B2 and FCC Ni or B19$'$ and FCC Ni phases. The fittings revealed the presence of $\sim 6 \pm 1$\% of the BCC Ni phase even in the stoichiometric NT50. This is in agreement with the slightly higher Ni content estimated from SEM-EDX (table \ref{tab:table2}). Best fits to the experimental spectra were obtained considering contributions from the near neighbor correlations consisting of Ti at $\sim$ 2.52 \AA, and at $\sim$ 2.60 \AA, Ni at $\sim$ 2.61 \AA~ and $\sim$ 2.84 \AA~  from the B19$'$ phase and a Ni--Ni scattering path at $\sim$ 2.44 \AA~ from the BCC Ni metal phase. The second scattering path from the BCC Ni metal phase (2.82 \AA) was masked by the Ni--Ni scattering path at 2.84 \AA~ from B19$'$ phase. On the other hand, the best fits to Ni EXAFS data in NT60 were obtained using Ni--Ti (2.61 \AA) and Ni--Ni (3.02 \AA) scattering paths from B2 phase and Ni--Ni scattering paths at 2.51 \AA~ and 3.54 \AA~ from the FCC Ni phase. The magnitude of the Fourier transform along with the best fitted curve for NT50 and NT60 alloys at 77 K are shown in Fig. \ref{fig:exp_NT50}. The best fitted curves to Ni EXAFS in all the alloys at 300 K and 77 K are presented in Fig.\ref{fig:exp_Ni}. The values of bond distances and the corresponding $\sigma^2$ are reported in the supplementary material \cite{Supplementary}. An increase in Ni--Ni bond length of the BCC Ni metal phase can be noticed with the increase in Ni concentration from a nominal value of $x$ = 0 to $x$ = 5. This increase in Ni--Ni bond distance is a clear indication of the evolution of Ni BCC phase to Ni FCC phase with an increase in excess Ni concentration.

\section{Discussion}

\begin{figure}[h]
\begin{center}
\includegraphics[width=\columnwidth]{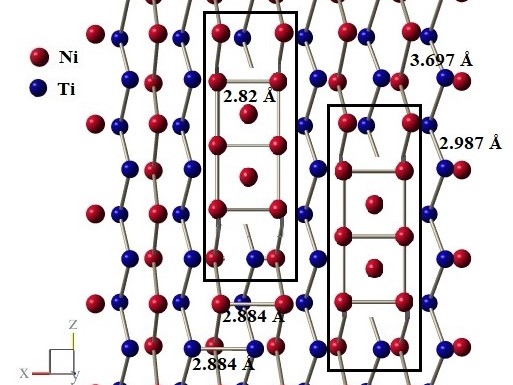}
\caption{Ball and stick model highlighting the structural distortion in B19$'$ structure due to presence of impurity Ni BCC clusters.}
\label{fig:structure}
\end{center}
\end{figure}

The local structural studies of Ni$_{50+x}$Ti$_{50-x}$ alloys enable us to understand the structural changes responsible for the development of strain glassy phase from martensite. The above analysis suggests that irrespective of the impurity dopant concentration, the Ti local structure remains consistent with the B19$'$ phase while the Ni EXAFS is made up of contributions from  the B19$'$ and BCC Ni phase at lower Ni excess concentration and to a mixture of B2 and FCC Ni metal phase at a higher excess Ni content. This discrepancy in local structures of the Ni and Ti is also manifested in the mean square radial displacement ($\sigma^2$) of Ti--Ni (Ti absorber) and Ni--Ti (Ni absorber) bonds. The $\sigma^2$ of Ni--Ti is higher ($\sim$ 0.03 \AA$^{2}$) than Ti--Ni ($\sim$ 0.01 \AA$^{2}$) bonds in all of the four alloy compositions \cite{Supplementary}, implying a larger disorder around the Ni in comparison with Ti.  This disorder around Ni is a result of defects, BCC Ni clusters which are formed in all Ni excess compositions. Such a possibility is depicted using a ball and stick model in Fig.\ref{fig:structure}. Further, the formation of defects would lead to increased scattering of free charge carriers and hence a higher resistivity. Indeed, the resistivity of the alloys increases with increasing $x$ in Ni$_{50+x}$Ti$_{50-x}$ \cite{Supplementary}.

With the excess Ni atoms progressively replacing the Ti atoms in Ni$_{50}$Ti$_{50}$,  the formation of Ni--Ni nearest neighbor pairs result in the growth of BCC Ni clusters (rectangular marked regions in Fig. \ref{fig:structure}). These nearest neighbor Ni--Ni pairs leverage the next nearest Ni--Ni bond lengths in the B19$'$ phase at $\sim$ 2.884 \AA~ to 2.82 \AA~  leading to its contraction and a structural defect. The presence of such structural defects forces the crystal structure to retain B2 symmetry even as the NiTi grain undergoes martensitic transformation. With further increase in Ni concentration, these defects grow in size increasing the possibility of formation of pure Ni clusters. At this point, the minimization of free energy demands a conversion of BCC Ni to FCC Ni. The strain caused by the presence of larger size defects is perhaps responsible for a change in the Ni local structure from B19$'$ to B2 type in Ni$_{60}$Ti$_{40}$ alloy. The formation of BCC Ni defects and the associated decrease in the in Ni--Ni bond length of B19$'$ phase explains the increase in FWHM of Bragg peaks and the decrease in the lattice constant with an increase in excess Ni concentration. The random distribution of Ni clusters breaks the long range ordering of the elastic strain vector, causing a martensitic composition transform to a strain glassy phase.

\section{Conclusions}
To summarize, we have studied the local structural changes in the Ni$_{50+x}$Ti$_{50-x}$ as the  alloys transit from martensite to strain glassy ground state with an increase in Ni impurity. EXAFS measurements at the Ni and Ti edge carried out at 300K and 77K show that due to doping of impurity Ni atoms for Ti, BCC Ni clusters, a defect phase segregates within the martensitic B19$'$ phase of the parent Ni$_{50}$Ti$_{50}$. With the increase in Ni content, nearest neighbor Ni--Ni correlations grow causing a structural change from BCC Ni to FCC Ni metal phase. All along these structural modifications, the core structure of the parent NiTi phase remains unaffected as evidenced from the unchanged Ti local structure throughout the doping region. The presence of BCC Ni clusters obstructs the propagation of long range ordering of the elastic strain vector, thereby leading to a transformation from a martensitic state to a strain glassy state.

\section*{Acknowledgements}
Financial assistance from the Science and Engineering Research Board, Govt. of India under the project SB/S2/CMP-0096/2013 is gratefully acknowledged. Department of Science and Technology, Govt. of  India and Elettra Sincrotrone, Trieste is thanked for the travel support and access within the framework of India Italy POC for the proposal 20185076. RN acknowledges the Council of Scientific and Industrial Research, Govt. of India for Senior Research fellowship.

\bibliographystyle{apsrev4-2}
\bibliography{Ref}

\end{document}